# EVIDENCE FOR R-MODE OSCILLATIONS IN SUPER-KAMIOKANDE SOLAR NEUTRINO DATA

P.A. STURROCK

Center for Space Science and Astrophysics, Varian 302, Stanford University, Stanford, California, 94305-4060, U.S.A.

(e-mail: sturrock@stanford.edu)

**Abstract.**

There has for some time been evidence of variability in radiochemical solar neutrino measurements, but this evidence has seemed suspect since the cerenkov experiments have not shown similar evidence of variability. The present re-analysis of Super-Kamiokande data shows strong evidence of r-mode oscillations. The frequencies of these oscillations correspond to a region with a sidereal rotation rate of 13.97 year$^{-1}$, not far from the value 13.88 year$^{-1}$ found some time ago in Homestake data, or 13.87 year$^{-1}$ found recently in GALLEX data. These estimates are incompatible with the rotation rate in the convection zone, but fully compatible with current estimates of the rotation rate in the radiative zone, including the solar core. These results are suggestive of variability originating in the core from fluctuating and asymmetric nuclear burning.

1. INTRODUCTION

We have in recent years searched for evidence of variability of the solar neutrino flux. The main (but not the only) approach has been to carry out a power-spectrum analysis and to search for peaks that might be identified with the frequency of solar rotation as seen from Earth (the so-called "synodic" frequency). Such an analysis of Homestake data (Davis and Cox, 1991) yielded a peak at 12.88 year$^{-1}$ (Sturrock, Walther, and Wheatland, 1977), corresponding to a sidereal rotation rate of 13.88 year$^{-1}$, which is compatible with rotation in the solar radiative zone (Schou et al., 1998; Garcia et al., 2008). However, subsequent analysis of GALLEX data (Sturrock, Scargle, Walther, and Wheatland, 1999; Sturrock and Scargle, 2001; Sturrock and Weber, 2002; Sturrock, Caldwell, and Scargle, 2006) yielded a prominent peak at 13.64 year$^{-1}$, which would correspond to a sidereal rotation rate of 14.64 year$^{-1}$, indicative of rotation in the convection zone. Since the convection zone contains magnetic structures, it seemed likely that such rotational modulation might be caused by the RSFP (Resonant Spin Flavor Precession; Akhmedov, 1988; Lim and Marciano, 1988) process. Detailed calculations of the possible role of RSFP in solar neutrino processes have been carried out by Chauhan and Pulido (2004), Chauhan, Pulido, and Raghavan (2005), and Balantekin and Volpe (2005).

However, a recent re-analysis of GALLEX data, using time-frequency analysis, has led to a different picture (Sturrock, 2008b). This analysis indicates that the principal peak in the power spectrum formed from GALLEX data is one at 13.87 year$^{-1}$, of which the peak at 13.64 year$^{-1}$ is merely an alias. (Power spectra formed from radiochemical data are unfortunately subject to strong aliasing, since the spacing between runs is typically a small multiple of one week.) This estimate of the frequency of rotational modulation is seen to be virtually identical to that previously derived from Homestake data (13.88 year$^{-1}$). This recent result directs attention to the radiative zone - including the core - rather than the convection zone. This is a surprising change, since one tends to associate any "activity" with the convection zone rather than the radiative zone (including the core), which are conventionally viewed as quiescent.

Concerning Super-Kamiokande data (Fukuda, 2001, 2002, 2003), our analyses have shown no evidence for rotational modulation, but they have pointed to possible evidence for an r-mode oscillation with frequency 9.43 year$^{-1}$ (Sturrock, 2004; Sturrock, Caldwell, Scargle, and Wheatland, 2005; Sturrock and Scargle, 2006; Sturrock, 2006). This suggests that we should look more closely into possible r-mode signatures in Super-Kamiokande data. That is the purpose of this article.

Since our analysis hinges on the possible role of r-modes in the solar interior, we briefly recapitulate the relevant formulas concerning r-modes in Section 2. We introduce two sets of frequencies: "S-type," which are the frequencies as they would be measured by an observer co-rotating with the Sun, and "E-type," which are the frequencies as they would be measured by an observer on Earth. The interaction of r-modes with structures (such as magnetic fields) embedded in the Sun would give rise to oscillations with S-type frequencies.

In order to make the case that r-modes may play an important role in solar dynamics, we briefly review the facts about the Rieger oscillation (Rieger et al., 1984) and related oscillations in Section 3, showing that these oscillations have frequencies that match very well S-type frequencies to be expected of r-modes that are excited in the tachocline (the region with normalized radius close to 0.7 where there is a sharp change between the rotation profile of the radiative zone and that of the convection zone) and interact with magnetic structures in that region or in a nearby region.

In Section 4, we review briefly evidence (in the form of S-type frequencies) for r-mode oscillations that have been found in our analyses of radiochemical solar neutrino data. In Section 5, we derive the power spectrum of Super-Kamiokande data, using a likelihood procedure that takes account of the beginning and end times and of the flux and error estimates for each data bin.

The new development in this article is contained in Section 6, where we search for E-type r-mode frequencies in the Super-Kamiokande power spectrum calculated in Section 5. We find evidence for a cluster of five related peaks. We estimate the significance of this cluster by using three procedures for combining power spectra. In Section 7, we obtain a more robust significance estimate by carrying out Monte Carlo simulations of the data. These calculations show that the cluster of peaks at E-type r-mode frequencies is highly unlikely to have occurred by chance. Analysis of the cluster leads to an estimate of the sidereal rotation frequency where the r-modes are excited. This is estimated to be 13.97 year$^{-1}$, not far from the frequency estimates previously found in our analyses of Homestake (13.88 year$^{-1}$) and GALLEX (13.87 year$^{-1}$) data. We discuss possible implications of these results in Section 8.

## 2. R-MODES

R-modes comprise a class of waves that can propagate in a rotating, uniform, inviscid sphere (Papaloizou and Pringle, 1978; Provost, Berthomieu, and Rocca, 1981; Saio, 1982). They are known in geophysics as "Rossby waves" (Pedlosky, 1987). An explanation of the physical process of r-modes is given by Saio (1982). Papaloizou and Pringle investigated the role of these waves in relation to short-period oscillations seen in cataclysmic variables. Wolff and Hickey (1987) pointed out the possible role of r-modes (in conjunction with g-modes) in solar physics. More recently, r-mode oscillations have been found to play a role in neutron stars, since they are unstable to the excitation of gravitational radiation (Anderson, 1998; Friedman and Morsink, 1998; Rezzolla, Lamb, Markovic, and Shapiro 2008a,b).

In a fluid sphere that rotates with frequency $\nu_R$, r-modes are retrograde with respect to the rotating fluid, with frequencies

$$\nu(l, m, S) = \frac{2m\nu_R}{l(l+1)}, \qquad (1)$$

where l and m are two of the three spherical-harmonic indices (the frequency is independent of n). The index l takes the values $l = 2, 3, ...$, and m takes the values $m = 1, ..., l$. We use the symbol "S" to denote that these are the frequencies that would be detected by an observer in the solar interior (rotating with the Sun).

As seen from Earth, and for waves rotating in the plane of the ecliptic, the frequencies (measured in cycles per year) are given by

$$v(l,m,E) = m(v_R - 1) - \frac{2mv_R}{l(l+1)}. \qquad (2)$$

It is significant that the frequency is independent of the radial index n, since we may construct an arbitrary radial profile from an appropriate combination of radial harmonic functions. In particular, we may if necessary consider a radial profile that is confined to a narrow radial band.

## 3. RIEGER-TYPE OSCILLATIONS

The tendency for gamma-ray flares to repeat with a period of about 154 days was discovered by Rieger et al. (1984). Investigators have reported that this periodicity is evident in general flare data and in daily sunspot area measurements (see, for instance, Bai & Cliver 1990 and references therein) and that there is evidence also for additional similar periodicities, notably one with period about 77 days (Bai 1992) and one with period about 51 days (Bai 1994; Kile & Cliver 1991).

Bai (2003), in a review of periodicities over sunspot cycles 19 - 23, drew special attention to oscillations with the following periods: 33.5 days, 51 days, 76 days, 84 days, 129 days, and 153 days, noting that four of the last five are close to being multiples (2, 3, 5, and 6) of 25.5 days, suggesting that this may be a "fundamental" period of the Sun. However, the periodicities of 33.5 days and 84 days do not fit this pattern.

The above six periodicities (in days) are listed in Column 8 of Table 1, and the corresponding frequencies (in year$^{-1}$) in Column 7. These periodicities are compared with frequencies (and periods) to be expected of r-modes (of type S or E) for various l, m, combinations, and for an assumed (sidereal) rotation frequency 14.32 year$^{-1}$, that provides the best fit to the 3-1-S, 3-2-S, and 3-3-S frequencies (given by Equation (1)). We see that the first four modes listed in that table do indeed have periods that are multiples of the period (25.43 days) corresponding to the assumed rotation frequency (14.32 year$^{-1}$). However, the fifth and sixth entries do not have periods that are integral multiples of 25.43 days. Hence the Rieger-type periods cannot all be interpreted as sub-harmonics of a "fundamental" periodicity.

On the other hand, we see that there is very good correspondence between all six of the frequencies and periods in columns 7 and 8 of Table 1 and the r-mode values listed in columns 4 and 5. Five of the six periodicities correspond to the frequencies of low-order r-modes in the solar frame, which interact with some structure (such as a localized magnetic field) that co-rotates with the Sun. The sixth periodicity (at 33.5 days) may be attributed directly to the effect of a traveling r-mode (l = 3, m = 1) wave.

In a recent article (Sturrock, 2008b), we have drawn attention to the fact that the Rieger-type oscillations may be attributed to r-mode oscillations in a spherical shell where the sidereal rotation frequency is at or near 14.32 year$^{-1}$. On examining estimates of the internal rotation rate derived from helioseismology (Schou 1998; Garcia et al. 2008), we find this shell to be located at or near the normalized radius 0.71 or 0.72, i.e. in or near the tachocline.

## 4. R-MODE OSCILLATIONS IN RADIOCHEMICAL SOLAR NEUTRINO DATA

In our analysis of Homestake data (Sturrock, Walther, and Wheatland, 1997), we drew attention to evidence for modulation of the solar neutrino flux at a frequency of 12.88 year$^{-1}$, which we attributed to rotation with the sidereal frequency 13.88 year$^{-1}$. We also drew attention to a periodicity with frequency 2.32 year$^{-1}$, which we suggested could be related to the Rieger periodicity. We see from Equation (1) that this frequency would correspond to a sidereal rotation frequency of 13.92 year$^{-1}$, close to the estimated rotational frequency of 13.88 year$^{-1}$.

In a recent article that focuses on GALLEX and GNO data (Sturrock, 2008b), we find evidence for rotational modulation for a sidereal frequency 13.87 year$^{-1}$, virtually identical to that found in our Homestake analysis. We also found evidence for periodicities close to some of the r-mode frequencies to be expected for that rotation frequency. We list in columns 7 and 8 of Table 2 the experimental frequencies and periods of oscillations identified in Homestake, GALLEX, and GNO power spectra. We list in columns 5 and 6 the theoretical frequencies and periods of r-mode oscillations for an assumed rotation rate of 13.87 year$^{-1}$. We see that there is good agreement between the frequencies identified in power spectra formed from radiochemical solar neutrino data and r-mode frequencies given by Equation (1), using the rotation rate determined from the radiochemical data.

A sidereal rotation frequency of 13.87 year$^{-1}$ is compatible with current estimates of the rotation rate of the radiative zone, including the core, although the rotation frequency of the core has not yet been well determined (Garcia et al. 2008).

## 5. POWER SPECTRUM FORMED FROM SUPER-KAMIOKANDE SOLAR NEUTRINO DATA

In previous articles (Sturrock, 2003, 2004; Sturrock, Caldwell, Scargle, and Wheatland, 2005; Sturrock and Scargle, 2006; Sturrock, 2006), we have examined power spectra derived from Super-Kamiokande data (Fukuda *et al.,* 2001, 2002, 2003). We have found no evidence of rotational modulation, but we did find a notable peak at 9.43 year$^{-1}$, which we suggested might be attributable to an r-mode with l = 2, m = 2. A similar peak (near 9.20 year$^{-1}$, but drifting in frequency) is found in the GNO power spectrum (Sturrock 2008b). Apart from this feature, we have found no evidence of periodicities of the "S" type, i.e. with frequencies given by Equation (1), in Super-Kamiokande data.

However, we saw in Table 1 that an r-mode oscillation of the "E" type (with l = 3, m = 1) shows up in the solar-activity analysis of Bai (2003). We also find some evidence for a similar periodicity in solar neutrino data: a peak at 10.52 year$^{-1}$ in Homestake data, one at 10.51 year$^{-1}$ in GALLEX data, and one at 10.57 year$^{-1}$ in SAGE data (Gavrin *et al.,* 2003). These are compatible with what is to be expected for an l = 3, m = 1, mode (E-type) for $v_R$ in the range 13.81 to 13.88 year$^{-1}$. For these reasons, it seems appropriate to examine the power spectrum formed from Super-Kamiokande data for evidence of r-mode frequencies of the E-type.

We have formed a power spectrum from Super-Kamiokande data (Fukuda *et al.,* 2001, 2002, 2003) by a likelihood procedure that takes account of the beginning and end time and of the flux and error estimates of each bin. (See, for instance, Sturrock 2003; Caldwell and Sturrock 2005.) The result is shown in Figure 1 for the frequency band 0 to 20 year$^{-1}$, and the top twenty peaks are listed in Table 3.

Since our investigations of Homestake and GALLEX data have yielded evidence for r-mode activity related to a sidereal rotation frequency close to 13.87 year$^{-1}$, we adopt 13 to 15 year$^{-1}$ as a reasonable search band for our investigation of Super-Kamiokande data. We see from Equation (1) that, for this band, and for l = 2, m = 2, an S-type r-mode frequency lies in the range 8.67 to 10 year$^{-1}$. As we have noted in previous articles (Sturrock, 2004; Sturrock, Caldwell, Scargle, and Wheatland, 2005), the peak at 9.43 year$^{-1}$, with power 11.24, is therefore a candidate for an r-mode periodicity. However, we do not find peaks for S-type r-mode frequencies corresponding to l = 2, m = 1, or to l = 3, m = 1, 2, or 3.

## 6. SEARCH FOR E-TYPE R-MODE OSCILLATIONS IN SUPER-KAMIOKANDE SOLAR NEUTRINO DATA

We now turn our attention to the E-type r-mode frequencies given by Equation (2). We find that, for m = 1, there appear to be peaks corresponding to l = 2, 3, 4, 5, and 6, as listed in Table 4. We now wish to determine whether there is a set of peaks that correspond to a common value of the rotation frequency. We can carry out this test by using any one of the three statistics we have introduced for the purpose of combining power spectra (Sturrock, Scargle, Walther, and Wheatland, 2005). These are the "Minimum Power Statistic" (MPS), the "Combined Power Statistic" (CPS), and the "Joint Power Statistic" (JPS). Each of these has the property that if the powers are independent and if each conforms to an exponential distribution, the statistic also conforms to an exponential distribution.

For the present case, in which we are combining five powers, the Minimum Power Statistic (MPS) is given by

$$U = 5 * Min(S_1,...,S_5). \quad (3)$$

This quantity is plotted, for the frequency band 5 – 20 year$^{-1}$, in Figure 2. The principal peak is found at 13.94 year$^{-1}$, and has the value $U_M = 10.40$.

The Combined Power Statistic (CPS) is a function of the sum of the powers. In terms of

$$Z = S_1 + ... + S_5, \quad (4)$$

the CPS is given by

$$G = Z - \log\left(1 + Z + \frac{1}{2!}Z^2 + \frac{1}{3!}Z^3 + \frac{1}{4!}Z^4\right). \quad (5)$$

This quantity is plotted, for the frequency band 5 – 20 year$^{-1}$, in Figure 3. We find that the second peak occurs at 13.97 year$^{-1}$, and has the value $G_M = 9.93$.

The Joint Power Statistic (JPS) is a function of the product of the powers. In terms of

$$Y = (S_1 * S_2 * ... * S_5)^{1/5}, \quad (6)$$

the JPS is given approximately by

$$J = \frac{4.9\, Y^2}{1.6 + Y}. \quad (7)$$

This quantity is plotted, for the frequency band 5 – 20 year$^{-1}$, in Figure 4. The biggest peak is found at 13.97 year$^{-1}$, and has the value $J_M = 11.48$.

We see that each analysis yields a prominent peak at or close to 13.97 year$^{-1}$. It is notable that this is very close to the value (13.88 year$^{-1}$) derived from our analysis of Homestake data (Sturrock, Walther, and Wheatland, 1997), and to the virtually identical value (13.87 year$^{-1}$) derived from our recent analysis of GALLEX data (Sturrock 2008b).

In Figure 5, we show the five peaks in the power spectrum that correspond to the r-mode E-type frequencies for m = 1, l = 2, 3, 4, 5, and 6, and $\nu_R = 13.97\,\text{year}^{-1}$. In Table 5, we list the expected frequencies, as given by Equation (2), and the actual peaks. We see that there is excellent agreement.

## 7. MONTE CARLO EVALUATION OF THE SIGNIFICANCE OF THE JOINT POWER STATISTIC

Of the three statistics considered in Section 6, the JPS is in some respects the most robust: the CPS does not require that each contribution be significant, and the MPS over-emphasizes the weakest peak. For this reason, we now focus on the Joint Power Statistic. In order to obtain a robust significance estimate of the results found in the previous section, we carry out Monte Carlo simulations. We generate 10,000 simulations of the Super-Kamiokande dataset by implementing the shuffle procedure (see, for instance, Bahcall and Press, 1991) as follows.

For the r'th data bin, we are given the start and end times $t_{s,r}$ and $t_{e,r}$, the flux estimate $g_r$, and the upper and lower error estimates $\sigma_{u,r}$ and $\sigma_{l,r}$. Our likelihood analysis uses the mean error estimate:

$$\sigma_{m,r} = \tfrac{1}{2}(\sigma_{u,r} + \sigma_{l,r}). \tag{8}$$

The duration of each bin is given by

$$D_r = t_{e,r} - t_{b,r}. \tag{9}$$

For each simulation, we retain the values of $D_r$, $g_r$, and $\sigma_{m,r}$, but we shuffle the values of $t_{e,r}$. We then compute the values of $t_{b,r}$ from $D_r$ and $t_{e,r}$. This procedure retains any relationship that may exist between the quantities $D_r$, $g_r$, and $\sigma_{m,r}$.

Since our earlier, independent, analyses of Homestake and GALLEX data led to estimates of the relevant internal rotation rate close to 14 year$^{-1}$, we adopt a search band of 13 to 15 year$^{-1}$ (410 – 475 nHz), which is wide enough to cover most of the most recent estimates of the rotation rate in the radiative zone and the convection zone (Garcia et al. 2008).

The result of this Monte Carlo analysis is shown in Figure 6. For only 27 of the 10,000 simulations do we find that the maximum value of the JPS over the frequency band 13 to 15 year$^{-1}$ is equal to or larger than the actual value 11.48. Hence, according to this test, the occurrence of this set of r-modes is significant at the 99.7% confidence level.

It is not unreasonable to narrow the search band to the range 13.77 to 13.97 year$^{-1}$, so as to examine only rotation frequencies as close to the value found in our radiochemical analyses (13.87 year$^{-1}$) as is the value found from our analysis of the actual Super-Kamiokande data (13.97 year$^{-1}$). The result of this Monte Carlo analysis is shown in Figure 7. For only 52 of the 100,000 simulations do we find that the maximum value of the JPS over the frequency band 13.77 to 13.97 year$^{-1}$ is equal to or larger than the actual value 11.48. Hence, according to this test, the occurrence of this set of r-modes is significant at the 99.95% confidence level.

## 8. DISCUSSION

Our earlier analyses of GALLEX data (Sturrock, Scargle, Walther, and Wheatland, 1999; Sturrock and Scargle, 2001; Sturrock and Weber, 2002; Sturrock, Caldwell, and Scargle, 2006) seemed to be indicative of variability originating in the convection zone, since the principal modulation frequency seemed to be in the band appropriate to that zone rather than the radiative zone. This result was suggestive of neutrino oscillations such as RSFP (Resonant Spin Flavor Precession; Akhmedov, 1988; Lim and Marciano, 1988) that can convert electron neutrinos into muon or tau neutrinos due to an interplay of a transition magnetic moment and a transverse magnetic field. This scenario has been investigated by Chauhan and Pulido (2004), Chauhan, Pulido, and Raghavan (2005), and Ballntekin and Volpe (2005). Caldwell (2007) has drawn attention to the possibility that this process may involve a sterile neutrino. However, according to the Balantekin-Volpe calculations, the RSFP process can lead to only small reductions of the neutrino flux, which would lead to correspondingly small depths of modulation. For neutrinos of energy less than 1 MeV (appropriate for gallium experiments), these calculations lead one to expect a depth of modulation smaller than one percent. On the other hand, our recent analysis of GALLEX data (Sturrock, 2008b) leads to an estimated depth of modulation in the range 80 - 100 percent. According to Balantekin (2008), such a large depth of modulation of low-energy neutrinos is incompatible with the RSFP hypothesis.

A further difficulty raised by our recent analysis of GALLEX data (which is consistent with our earlier analysis of Homestake data) is that the modulation frequency points to processes in the radiative zone rather than the convection zone. Dynamo theories of the solar magnetic field (see, for instance, Krause, Raedler, and Ruediger, 1993) point to processes in or near the tachocline, generating field in the convection zone only. This poses a further problem for the hypothesis that the solar neutrino flux is modulated by processes - such as RSFP - in the convection zone.

The results of the present analysis of Super-Kamiokande data also point to modulation that occurs in the radiative zone (possibly in the core) rather than in the convection zone. This new result, together with difficulties with the RSFP scenario noted above, suggests that we look for an alternative interpretation of variability of the solar neutrino flux. The most obvious possibility is that variability may originate in the nuclear-burning process itself.

In constructing solar models, it is generally assumed that the internal structure is, to good approximation, steady and spherically symmetric. This may be true of the radiative zone, which is thought to be a passive part of the Sun, but it is known not to be true of the convection zone, which involves energy conversion from radiation to convection, and it may well not be true of the core since nuclear burning generates energy that could have dynamical consequences. If the core is in fact dynamic rather than static, it could well be asymmetric. Asymmetry in the nuclear burning, together with the matter-induced MSW neutrino oscillations (Wolfenstein, 1978; Mikheyev and Smirnov, 1985), could explain variability of the neutrino flux with a frequency characteristic of the rotation rate of the core. Current estimates of the internal rotation rate (below the tachocline) are consistent with more-or-less rigid rotation with a frequency in the range 430 to 450 nHz, i.e. 13.6 to 14.2 year$^{-1}$ (Garcia et al. 2008). Hence our findings of modulations corresponding to rotation rates in the range 13.87 to 13.97 year$^{-1}$ are clearly consistent with the hypothesis that this variability may originate as rotational modulation in the core.

We saw in Sections 2 and 3 that Rieger-type oscillations are most probably a manifestation of r-mode oscillations that are excited in the vicinity of the tachocline. This is a region of strong velocity shear, which is known, from plasma physics, to be source of free energy for driving an instability. Since nuclear reactions are also a source of energy, it is not unreasonable to suppose that this scenario also may lead to an instability involving r-modes. This proposal is reminiscent of the fact that an interplay of gravitational radiation and r-modes leads to the development of r-modes in neutron stars (Anderson, 1998; Friedman and Morsink, 1998; Rezzolla, Lamb, Markovic, and Shapiro 2008a,b). Hence nuclear burning may lead both to the asymmetry that leads to rotational modulation, and to the excitation of r-modes, that we have found to be evident in Super-Kamiokande data.

A further consequence of hypothetical variability in nuclear burning is that the apparent bimodality in GALLEX flux measurements (Sturrock and Scargle, 2001; Sturrock, 2008a) is no longer an anomaly. The solar luminosity sets a firm upper limit to the time-averaged neutrino generation rate, but it does not set a limit to the instantaneous rate.

It should be noted that if we entertain the possibility that the core may vary on a timescale of months, we should logically be willing to also entertain the possibility that it may vary on a timescale of years. This has the awkward consequence that although the solar luminosity informs us of the mean neutrino generation rate, it does not inform us of the generation rate during the period of operation of any particular experiment. Most discussions and analyses of the solar neutrino flux have been based on the assumption that the flux is constant: based on this assumption, one can derive an estimate of the flux generated in the core from knowledge of the solar luminosity. On combining this estimate with values of the neutrino flux derived from experiments, one can determine the factor by which the flux is reduced in transit from the core;

this reduction has been attributed to neutrino oscillations. If nuclear burning is variable, this chain of argument must be re-evaluated.

## Acknowledgements

Thanks are due to Bala Balantekin, David Caldwell, Alexander Kosovichev, Joao Pulido, Jeffrey Scargle, Guenther Walther, Michael Wheatland, and Steven Yellin for helpful discussions related to this work, which was supported by NSF Grant AST-0607572.

**Tables**

Table 1. Rieger-Type Periodicities for an assumed sidereal rotation frequency $v_R = 14.32\,year^{-1}$.

| l | m | S/E | v(exp) (yr$^{-1}$) | P(exp) (days) | P/P$_R$ | v(Bai) (yr$^{-1}$) | P(Bai) (days) |
|---|---|-----|--------------------|---------------|---------|--------------------|---------------|
| 3 | 1 | S | 2.39 | 152.6 | 6 | 2.39 | 153 |
| 3 | 2 | S | 4.79 | 76.3 | 3 | 4.80 | 76 |
| 3 | 3 | S | 7.18 | 50.9 | 2 | 7.16 | 51 |
| 4 | 2 | S | 2.87 | 127.2 | 5 | 2.83 | 129 |
| 4 | 3 | S | 4.31 | 84.8 | 3.33 | 4.35 | 84 |
| 3 | 1 | E | 10.97 | 33.3 | 1.31 | 10.90 | 33.5 |

Table 2. R-mode frequencies, as expected for an assumed rotational frequency $v_R = 13.87\,year^{-1}$, and as possibly identified in radiochemical solar neutrino data.

| Experiment | l | m | S/E | v(exp) (yr$^{-1}$) | P(exp) (days) | v(actual) (yr$^{-1}$) | P(actual) (days) |
|------------|---|---|-----|--------------------|---------------|------------------------|------------------|
| Homestake | 3 | 1 | S | 2.31 | 157.9 | 2.32 | 157 |
| GALLEX | 3 | 2 | S | 4.63 | 78.9 | 4.54 | 80.5 |
| GALLEX | 3 | 3 | S | 6.94 | 52.6 | 6.93 | 52.7 |
| GNO | 4 | 2 | S | 9.25 | 39.5 | 9.20 | 39.7 |

Table 3. Top twenty peaks in the range $0-20\ year^{-1}$ in the power spectrum formed from Super-Kamiokande data by a likelihood procedure.

| Frequency (year$^{-1}$) | Power |
|---|---|
| 9.43 | 11.24 |
| 12.31 | 6.01 |
| 8.30 | 5.26 |
| 12.69 | 4.44 |
| 15.72 | 4.1 |
| 0.36 | 4.03 |
| 4.43 | 3.68 |
| 18.10 | 3.67 |
| 18.76 | 3.63 |
| 10.69 | 3.45 |
| 8.98 | 3.26 |
| 11.29 | 3.18 |
| 17.58 | 3.11 |
| 5.90 | 2.72 |
| 0.80 | 2.68 |
| 11.57 | 2.66 |
| 14.87 | 2.63 |
| 2.96 | 2.54 |
| 8.75 | 2.49 |
| 2.31 | 2.32 |

Table 4. Candidate peaks in the Super-Kamiokande power spectrum for E-type r-mode frequencies corresponding to m = 1, l = 2, 3, 4, 5, and 6.

| l | Search band (yr$^{-1}$) | Frequency (yr$^{-1}$) | Power |
|---|---|---|---|
| 1 | | | |
| 2 | 7.67 - 9.00 | 8.30 | 5.26 |
| 3 | 9.83 - 11.50 | 10.69 | 3.45 |
| 4 | 10.70 - 12.50 | 11.57 | 2.66 |
| 5 | 11.13 - 13.00 | 12.01 | 2.08 |
| 6 | 11.38 - 13.29 | 12.31 | 6.01 |

Table 5. Comparison of actual peaks in the Super-Kamiokande power spectrum and expected frequencies of E-type r-modes corresponding to m = 1, l = 2, 3, 4, 5, and 6, with $v_R = 13.97\,\text{year}^{-1}$.

| l | Expected Frequency (yr$^{-1}$) | Actual Frequency (yr$^{-1}$) | Power | DOM % |
|---|---|---|---|---|
| 2 | 8.31 | 8.30 | 5.26 | 4.7 |
| 3 | 10.64 | 10.69 | 3.45 | 3.6 |
| 4 | 11.57 | 11.57 | 2.66 | 3.4 |
| 5 | 12.04 | 12.01 | 2.08 | 3.0 |
| 6 | 12.30 | 12.31 | 6.01 | 5.1 |

**Figures**

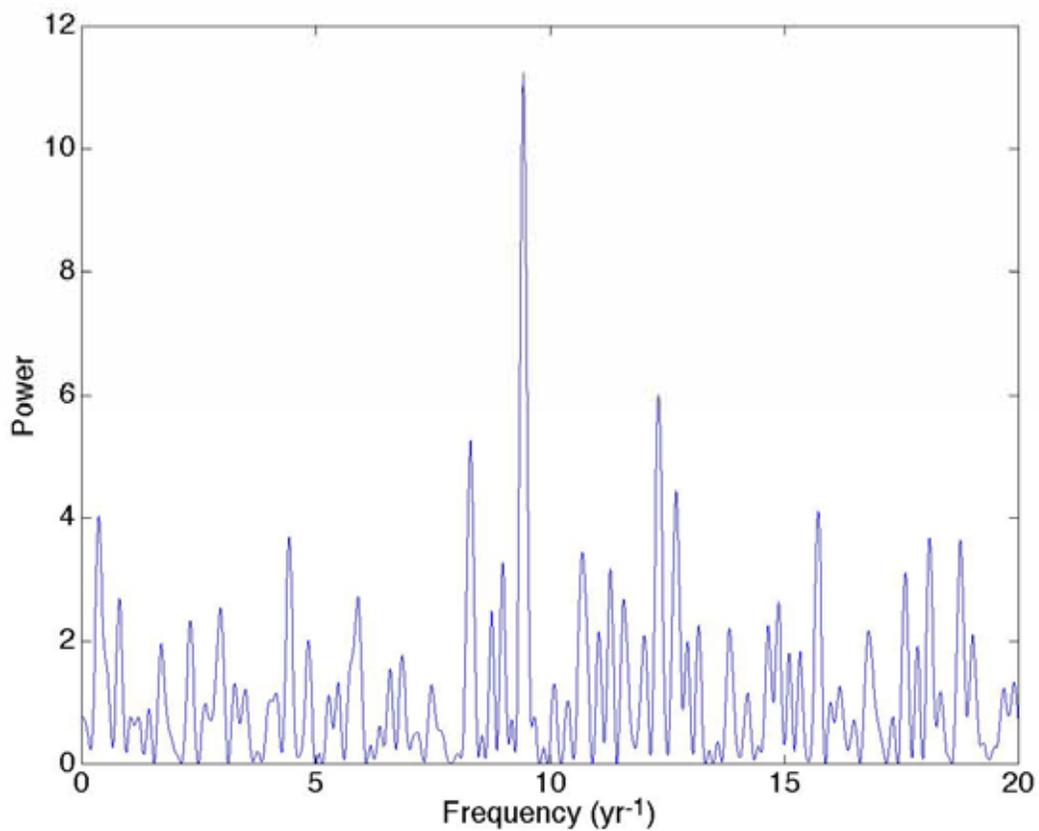

Figure 1. Power spectrum formed from Super-Kamiokande data, using a likelihood procedure.

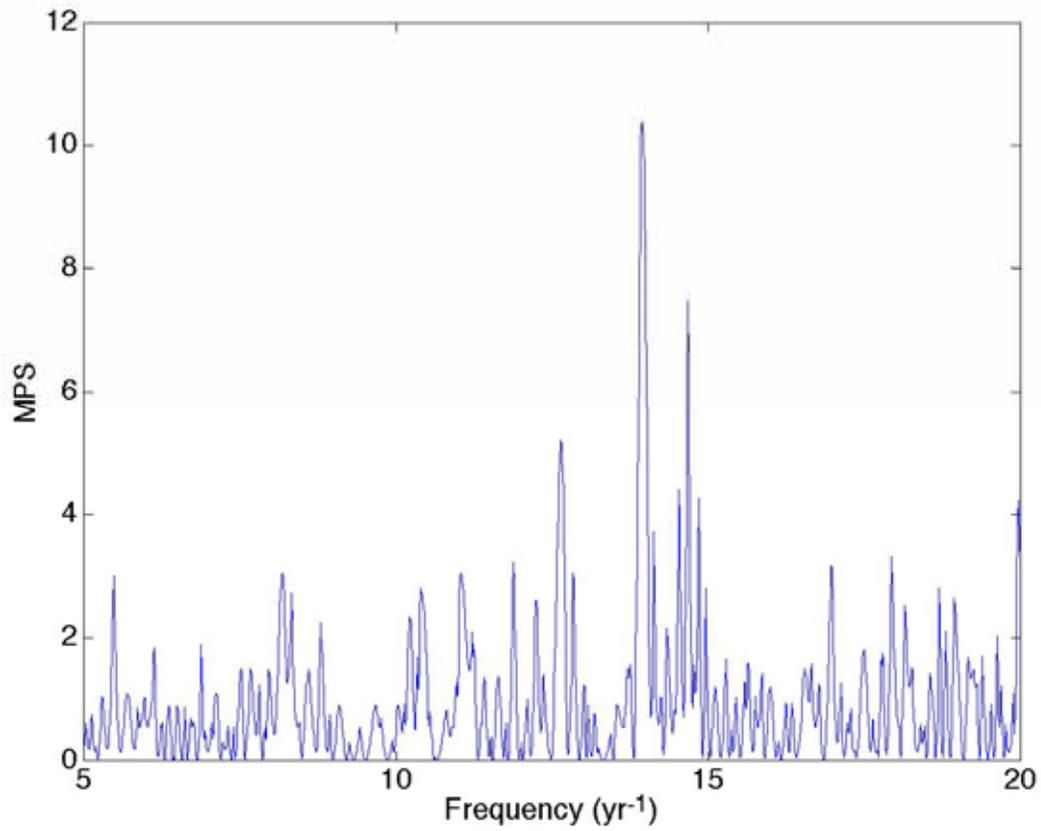

Figure 2. The Minimum Power Statistic formed from the powers of the m = 1, l = 2, 3, 4, 5, and 6 E-type r-mode frequencies.

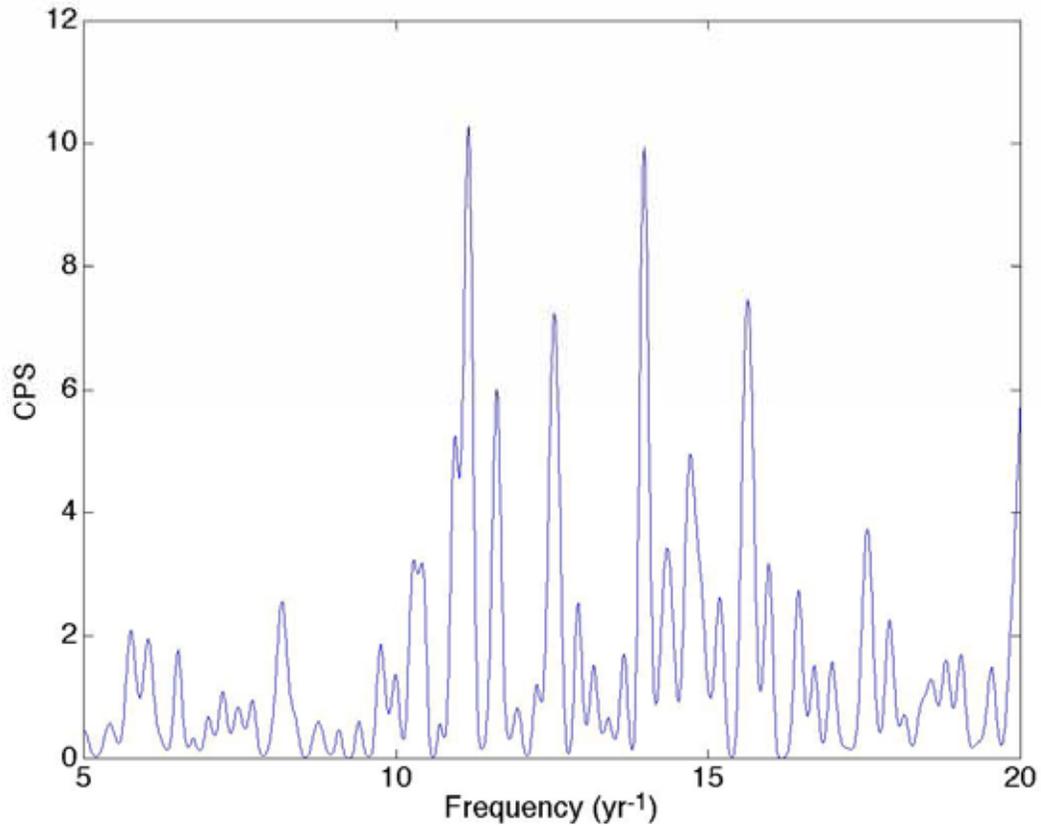

Figure 3. Combined Power Statistic formed from the powers of the m = 1, l = 2, 3, 4, 5, and 6 E-type r-mode frequencies.

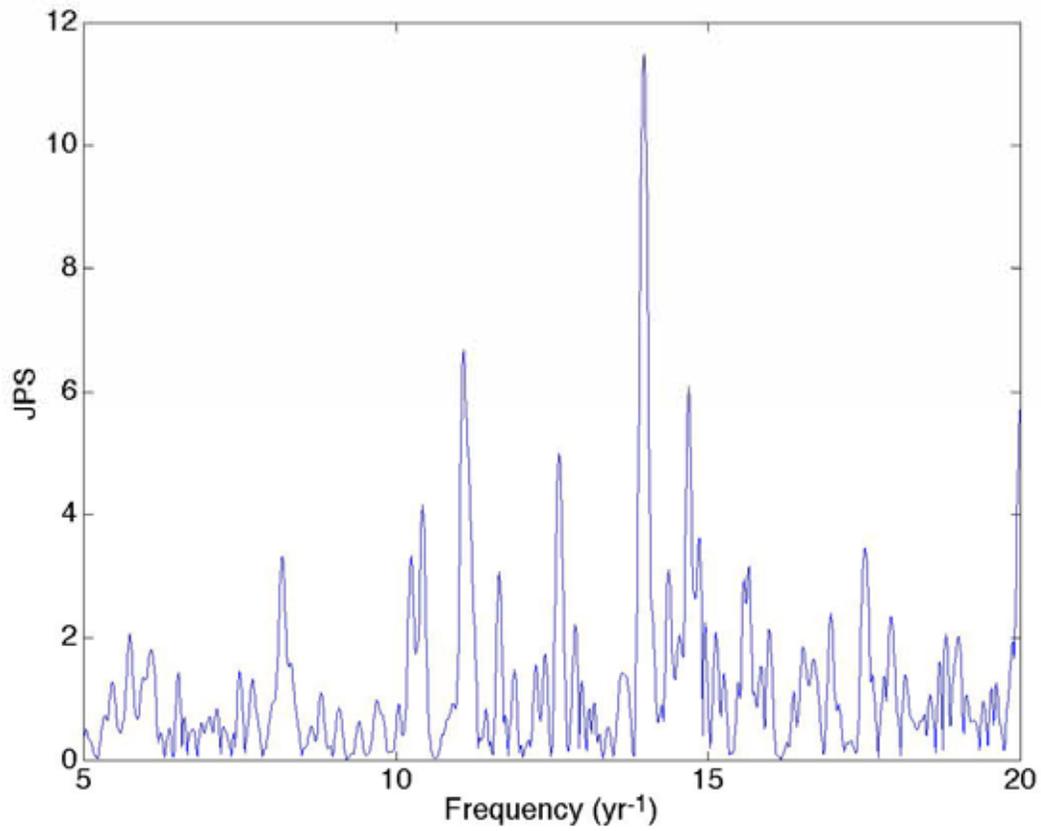

Figure 4. Joint Power Statistic formed from the powers of the m = 1, l = 2, 3, 4, 5, and 6 E-type r-mode frequencies.

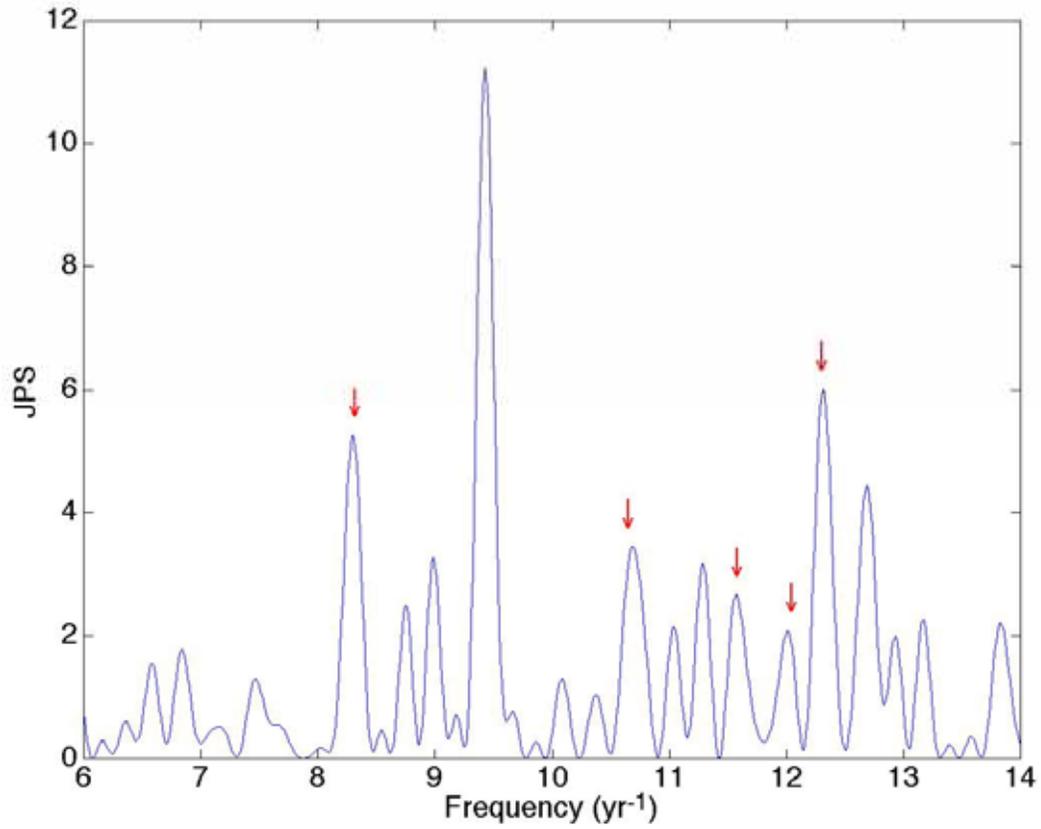

Figure 5. Excerpt of power spectrum, showing the expected locations of the peaks (for the m = 1, l = 2, 3, 4, 5, and 6 E-type r-mode frequencies for $v_R = 13.97\,\text{year}^{-1}$) in relation to the actual peaks.

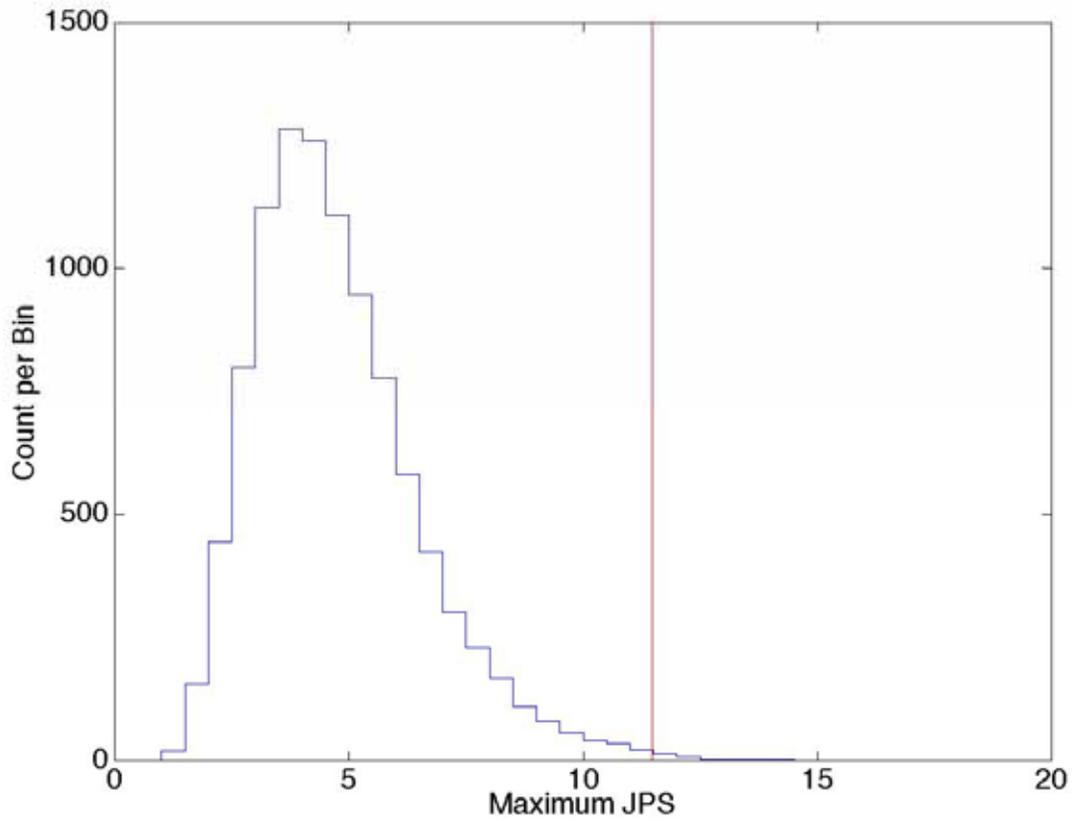

Figure 6. Histogram of peak values of the Joint Power Statistic formed from 10,000 simulations over the frequency range 12 – 14 year$^{-1}$. Only 27 simulations have values as large as or larger than the actual value (11.48) indicated by the vertical red line.

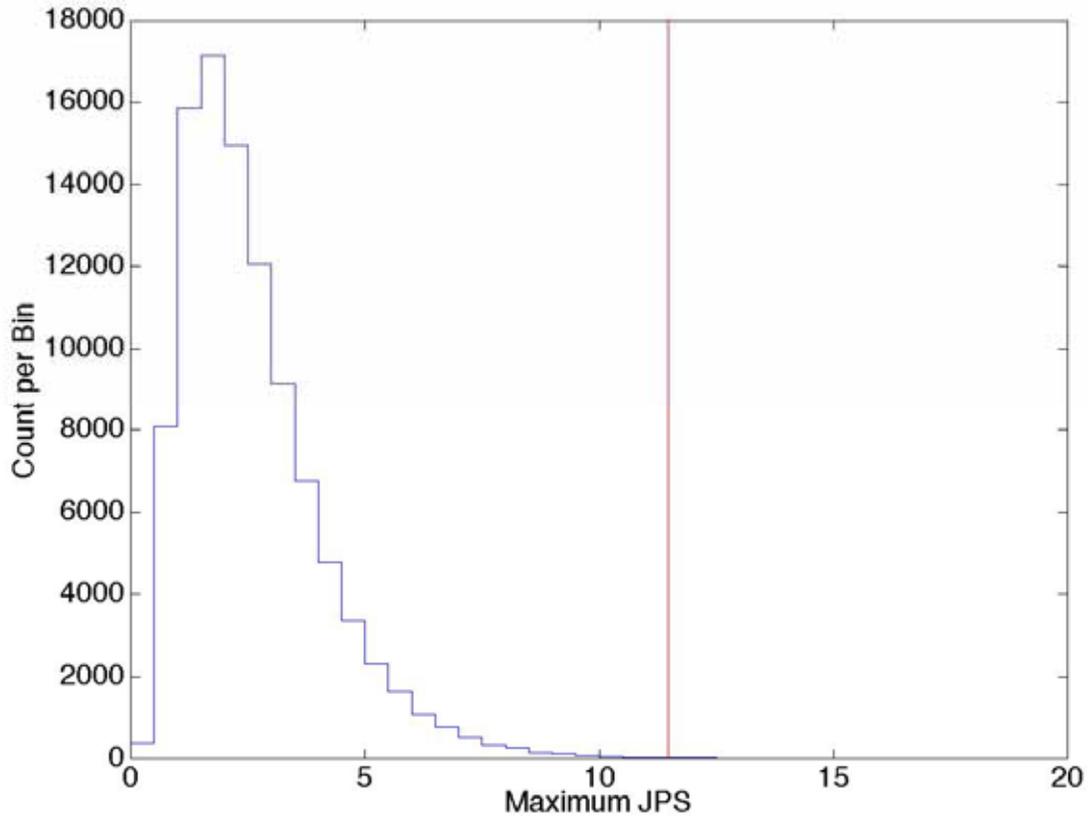

Figure 7. Histogram of peak values of the Joint Power Statistic formed from 100,000 simulations over the frequency range 13.77 – 13.97 year$^{-1}$. Only 52 simulations have values as large as or larger than the actual value (11.48) indicated by the vertical red line.